\documentclass[useAMS,usenatbib]{mn2e}

\usepackage{amsmath, amsfonts,amssymb,theorem,graphics,epsf }
\begin{document}
\def\lsun{\ifmmode{L_\odot}\else$L_\odot$\fi}
\def\msun{\ifmmode{M_\odot}\else$M_\odot$\fi}
\def\rsun{{\mathrm{R}}_\odot}
\def\Teff{{T_{\mathit{eff}}}}
\newcommand{\gsimeq}{\mbox{$\, \stackrel{\scriptstyle >}{\scriptstyle
\sim}\,$}}
\def\gradi{\ifmmode{^\circ}\else$^\circ$\fi}
\def\reference{\parskip 0pt\par\noindent\hangindent 0.5 truecm}

\title[The initial-final mass relationship] {The open cluster initial-final
mass relationship and the high-mass tail of the white dwarf distribution}
\author[Ferrario et al.]  {Lilia Ferrario $^{1}$, Dayal Wickramasinghe$^{1}$, 
James
Liebert$^{2}$ and Kurtis A. Williams$^{2}$\\ \\ $^{1}$The Australian National
University, ACT 0200, Australia\\ $^{2}$Steward Observatory, University of
Arizona, Tucson, AZ 85726, USA}

\date{}
\pagerange{\pageref{firstpage}--\pageref{lastpage}} \pubyear{2005}
\maketitle
\label{firstpage}

\begin{abstract}
Recent studies of white dwarfs in open clusters have provided new constraints
on the initial - final mass relationship (IFMR) for main sequence stars with
masses in the range $2.5 - 6.5\msun$. We re-evaluate the ensemble of data that
determines the IFMR and argue that the IFMR can be characterised by a mean
initial-final mass relationship about which there is an intrinsic scatter. We
investigate the consequences of the IFMR for the observed mass distribution of
field white dwarfs using population synthesis calculations. We show that while
a linear IFMR predicts a mass distribution that is in reasonable agreement
with the recent results from the PG survey, the data are better fitted by an
IFMR with some curvature. Our calculations indicate that a significant ($\sim
28$) percentage of white dwarfs originating from single star evolution have
masses in excess of $\sim 0.8\msun$ obviating the necessity for postulating
the existence of a dominant population of high-mass white dwarfs that arise
from binary star mergers.
\end{abstract}


\section{Introduction}

A knowledge of the Initial - Final mass relationship (IFMR) of stars is of
fundamental importance in many astrophysical contexts and plays a key role in
our understanding of chemical enrichment and the efficiency of star formation
in galaxies. The low and intermediate mass end of this relationship can be
probed by studying the properties of white dwarfs in open clusters.

The IFMR can also be investigated by looking at the mass distribution of field
white dwarfs. Here, we are dealing with white dwarfs which would have had an
origin in stellar clusters with a variety of parameters during the history of
the galactic disc. Earlier studies (e.g. Koester \& Weidemann 1980) showed
that the predominant characteristic of the mass distribution, a sharp peak at
$M_f = 0.57\pm .01$, could be explained by the steep initial mass function and
the finite age of the galactic disc. This peak corresponded very closely to
the current turn-off mass of the stellar population in the galactic disc.  The
nature of the mass distribution above and below this peak clearly contained
important additional information on the progenitors of the white dwarfs, but
this could not be fully exploited at that time because of lack of data.

Recently, detailed studies have been carried out on large samples of white
dwarfs in surveys such as the PG and the SDSS.  With these new data, the
relative importance of massive ($M \ge 0.8\msun$) white dwarfs only became
apparent when attempts were made to correct observations based on magnitude
limited surveys for the mass dependence of the intrinsic luminosity of the
white dwarfs.  The first such results for the PG survey were presented for the
magnetic white dwarfs by Liebert Bergeron \& Holberg (2003).  They showed that
when corrections were made to allow for the fact that the more massive
magnetic white dwarfs had smaller radii, and hence were less luminous at a
given effective temperature, the space density of high-field magnetic white
dwarfs (HFMWDs) more than doubled from the previously estimated value of $\sim
4$ percent to $\sim 10$ percent.  Wickramasinghe \& Ferrario (2005, hereafter
WF05) used population synthesis calculations to show that these observations
could be explained if a significant fraction of the HFMWDs had massive
progenitors.  The mass distribution of the entire DA white dwarf population
(magnetic and non-magnetic) in the PG sample has only recently been
established by Liebert, Bergeron \& Holberg (2005a). These data, when corrected
to an equivalent volume limited sample, showed the presence of a significant
high-mass tail with $\sim 30$ per cent of DA white dwarfs having masses in
excess of $0.8\msun$.  The details of this mass distribution must hold
important clues on the nature of the mean IFMR appropriate to stars in the
galactic disc.

In this paper we re-evaluate the open cluster data that are currently being
used to define an IFMR and characterise this relationship. We show that the
nature of this relationship has observable consequences for the mass 
distribution 
of field white dwarfs and show how the latter can be used to place constraints 
on the IFMR.

\section{Analysis of Cluster Data}

We use data for the clusters listed below. Only white dwarfs which are
probable cluster members are included from the published data sets. The masses
of the progenitor stars are calculated using the nuclear lifetimes from the
tables of Girardi et al. (2002) for $Z= 0.008, 0.019, 0.03$. The white dwarf
mass and cooling age are obtained from the Fontaine, Brassard, \& Bergeron
(2001) evolutionary models which incorporate improvements in the equation of
state and conductive opacities.

\emph{Hyades and Praesepe} --- The data points are from Claver et al. (2001)
supplemented with recent results on two new stars in Praesepe (Dobbie et
al. 2005). Perryman et al. (1998) estimate an age $625^{+50}_{-50}$ Myr with
the errors given by von Hippel (2005) for the Hyades.  The isochrone fit in
Claver et al. (2001) led these authors to adopt the same age for Praesepe.
Both have been linked to the same Hyades Supercluster and have a metallicity
of [Fe/H] = $0.11-0.15$ (King \& Hiltgen 1996; Cayrel, Cayrel de Strobel \&
Campbell 1985; Boesgaard \& Friel 1990). We have used $Z=0.019$.

\emph{Pleiades}--- The age of the Pleiades is $125^{+25}_{-25}$ Myr from isochrones
calculated with moderate convective core overshooting (Mazzei \& Pigatto 1989;
Meynet, Mermilliod \& Maeder 1993; Ventura et al. 1998).  This is in agreement
with the age determined from the luminosity at which lithium is detected in
the substellar objects (Stauffer, Schulz, \& Kirkpatrick 1998). Since the
metallicity of the Pleiades does not differ more than $\pm 0.1$ dex from solar
(e.g. King \& Soderblow 2000), the Girardi et al. (2002) $Z=0.019$ table was
used. We have adopted the $\Teff$ obtained from the spectral fit, but have
modified the gravity to give the gravitational redshift mass of $1.02
\msun$. 

\emph{NGC2168 (M35)} --- The cluster data are from Williams et al. (2004)
supplemented by new data on LAWDS 2 and LAWDS 27. We adopt an age $150\pm 60$
Myr given by von Hippel (2005) and used by Williams et al. (2004).  A separate
study by Steinhauer (2003) gives an age of $160\pm 20$ Myr. Sung \& Bessell
(1999) obtain a value of [Fe/H] = -0.3 while Twarog et al. (1997) give [Fe/H]
= -0.16. In our studies, we use $Z=0.008$.

\emph{NGC2516} --- The cluster data are from Koester \& Reimers (1996). Meynet
et al. 1993 estimate an age $158^{+20}_{-20}$ Myr. Sung et al. (2002) give an
abundance [Fe/H]= $-0.10\pm 0.04$. We have used $Z=0.019$.

\emph{NGC3532} --- The cluster data are from Koester \& Reimers (1993). The
$\log g$ and $\Teff$ used by us are the mean values of the observations in
Koester \& Reimers (1996). This sparse cluster has an estimated age of
$300^{+150}_{-150}$ Myr. Twarog et al. (1997) give a metallicity [Fe/H] =
$-0.022$. Thus, we have used the $Z= 0.019$ table.

\emph{NGC2099 (M37)} --- We have used the results of Kalirai et al. (2005).
This set of data are of somewhat lower quality because these stars are much
fainter than all other stars considered here ($V\sim 22-23.6$) and as a
result. the coverage of the high Balmer lines which are most sensitive to
gravity (and stellar mass) was generally of poorer quality. In addition, the
stars WD 1, 15, 17, 24 are excluded because they are at the wrong distance to
be members of this cluster.  We note also that WD 15 and 17 have cooling ages
that are much larger than any plausible cluster age.  WD6 and WD21 are also
excluded because the gravity was determined assuming cluster membership,
rather than from the spectrum.  We have adopted the age of $520^{+80}_{-80}$
and solar metallicity (Twarog et al. 1997).

\emph{Sirius} --- We include new data by Liebert et al. (2005b) who used the
well determined luminosity and radius of Sirius A and stellar evolution models
with a solar metallicity to accurately constrain the age for this system to
$238^{+13}_{-13}$.

Error bars for the initial and final masses were determined using a Monte
Carlo simulation.  1000 white dwarfs were created with Gaussian distributions
of $\Teff$ and $\log g$ with $1\sigma$ equivalent to the published errors on
these quantities. This makes the implicit assumption that the derived $\Teff$
and $\log g$ are independent. The $M_i$ and $M_f$ were then calculated for
each simulated sample, and the 68 percent confidence limits. These results are
presented together with the white dwarf cooling ages $\tau$ in Table 1. Two
sets of results are given for the errors. The first assumes the best estimate
for the age of a cluster so that the errors are those from the white dwarf
spectral observations.  The second gives the systematic errors that result
purely from the uncertainties on the cluster ages. The total error bars are
obtained by adding these errors in quadrature as shown in Figure 1.

\setcounter{table}{0}
\begin{table*}
 \begin{minipage}{168mm}
 \caption{Cluster data.}
 \label{tab1}
 \begin{tabular}{@{}cllllllll}
\hline
Cluster& White Dwarf        & $\Teff\pm d\Teff$& $\log g\pm d\log g$&$M_f\pm dM_f$ &$\log\tau\pm d\log\tau$&  $M_i$  & $dM_i$               &  $dM_i$             \\
                            &                &                &                    &                    &  &         & (random)             & (systematic)        \\
\hline
Pleiades&  LB1497           & $31660\pm 350$ & $8.64$         & $ 1.023\pm 0.026 $ & $ 7.781\pm 0.062 $ &  $ 6.542 $ & $_{-0.346}^{+0.458}$ & $_{-0.866}^{+1.614}$ \\ 
Haydes &  0352+098          & $16629\pm 350$ & $8.16\pm 0.05$ & $ 0.719\pm 0.030 $ & $ 8.270\pm 0.046 $ &  $ 3.094 $ & $_{-0.045}^{+0.052}$ & $_{-0.114}^{+0.134}$ \\ 
\&     &  0406+169          & $15180\pm 350$ & $8.30\pm 0.05$ & $ 0.806\pm 0.031 $ & $ 8.488\pm 0.046 $ &  $ 3.465 $ & $_{-0.109}^{+0.147}$ & $_{-0.172}^{+0.225}$ \\ 
Praesepe& 0421+162          & $19570\pm 350$ & $8.09\pm 0.05$ & $ 0.680\pm 0.031 $ & $ 7.970\pm 0.055 $ &  $ 2.892 $ & $_{-0.020}^{+0.023}$ & $_{-0.090}^{+0.103}$ \\ 
       &  0425+168          & $24420\pm 350$ & $8.11\pm 0.05$ & $ 0.705\pm 0.031 $ & $ 7.549\pm 0.077 $ &  $ 2.789 $ & $_{-0.011}^{+0.010}$ & $_{-0.079}^{+0.088}$ \\ 
       &  0431+125          & $21340\pm 350$ & $8.04\pm 0.05$ & $ 0.652\pm 0.032 $ & $ 7.752\pm 0.068 $ &  $ 2.825 $ & $_{-0.014}^{+0.017}$ & $_{-0.083}^{+0.093}$ \\ 
       &  0438+108          & $27390\pm 350$ & $8.07\pm 0.05$ & $ 0.684\pm 0.031 $ & $ 7.203\pm 0.060 $ &  $ 2.757 $ & $_{-0.003}^{+0.004}$ & $_{-0.076}^{+0.084}$ \\ 
       &  0437+138          & $15335\pm 350$ & $8.26\pm 0.05$ & $ 0.782\pm 0.033 $ & $ 8.447\pm 0.047 $ &  $ 3.365 $ & $_{-0.089}^{+0.117}$ & $_{-0.156}^{+0.192}$ \\ 
       &  0836+197          & $21949\pm 350$ & $8.45\pm 0.05$ & $ 0.909\pm 0.030 $ & $ 8.136\pm 0.049 $ &  $ 2.981 $ & $_{-0.031}^{+0.035}$ & $_{-0.101}^{+0.115}$ \\ 
       &  0836+201          & $16629\pm 350$ & $8.01\pm 0.05$ & $ 0.620\pm 0.031 $ & $ 8.154\pm 0.050 $ &  $ 2.993 $ & $_{-0.028}^{+0.046}$ & $_{-0.102}^{+0.117}$ \\ 
       &  0836+199          & $14060\pm 630$ & $8.34\pm 0.06$ & $ 0.831\pm 0.037 $ & $ 8.612\pm 0.069 $ &  $ 3.997 $ & $_{-0.350}^{+0.652}$ & $_{-0.297}^{+0.428}$ \\ 
       &  0837+199          & $17098\pm 350$ & $8.32\pm 0.05$ & $ 0.819\pm 0.032 $ & $ 8.351\pm 0.048 $ &  $ 3.194 $ & $_{-0.062}^{+0.081}$ & $_{-0.129}^{+0.153}$ \\ 
       &  0837+218          & $17845\pm 560$ & $8.48\pm 0.075$& $ 0.918\pm 0.045 $ & $ 8.422\pm 0.073 $ &  $ 3.315 $ & $_{-0.124}^{+0.168}$ & $_{-0.148}^{+0.178}$ \\ 
       &  0837+185          & $14170\pm 1485$& $8.46\pm 0.15$ & $ 0.900\pm 0.088 $ & $ 8.683\pm 0.171 $ &  $ 4.695 $ & $_{-1.131}^{+9.990}$ & $_{-0.523}^{+0.901}$ \\ 
       &  0840+200          & $14178\pm 350$ & $8.23\pm 0.05$ & $ 0.761\pm 0.033 $ & $ 8.522\pm 0.045 $ &  $ 3.572 $ & $_{-0.129}^{+0.186}$ & $_{-0.196}^{+0.255}$ \\ 
NGC2516&  2516-1            & $28170\pm 310$ & $8.48\pm 0.17$ & $ 0.931\pm 0.098 $ & $ 7.760\pm 0.230 $ &  $ 5.411 $ & $_{-0.500}^{+1.017}$ & $_{-0.386}^{+0.524}$ \\ 
       &  2516-2            & $34200\pm 610$ & $8.60\pm 0.11$ & $ 1.004\pm 0.058 $ & $ 7.621\pm 0.179 $ &  $ 5.096 $ & $_{-0.239}^{+0.406}$ & $_{-0.311}^{+0.415}$ \\ 
       &  2516-3            & $26870\pm 330$ & $8.55\pm 0.07$ & $ 0.969\pm 0.039 $ & $ 7.922\pm 0.079 $ &  $ 6.141 $ & $_{-0.427}^{+0.692}$ & $_{-0.587}^{+0.924}$ \\ 
       &  2516-5            & $30760\pm 420$ & $8.70\pm 0.12$ & $ 1.054\pm 0.063 $ & $ 7.883\pm 0.136 $ &  $ 5.902 $ & $_{-0.524}^{+0.986}$ & $_{-0.514}^{+0.786}$ \\ 
NGC3532&  3532-8            & $23367\pm 1065$&$7.713\pm 0.148$& $ 0.481\pm 0.065 $ & $ 7.324\pm 0.098 $ &  $ 3.634 $ & $_{-0.014}^{+0.028}$ & $_{-0.514}^{+1.259}$ \\ 
       &  3532-9            & $29800\pm 616$ &$7.827\pm 0.229$& $ 0.551\pm 0.111 $ & $ 6.960\pm 0.107 $ &  $ 3.578 $ & $_{-0.002}^{+0.008}$ & $_{-0.488}^{+1.142}$ \\ 
       &  3532-10           & $19267\pm 974$ &$8.143\pm 0.266$& $ 0.712\pm 0.158 $ & $ 8.043\pm 0.244 $ &  $ 4.200 $ & $_{-0.379}^{+0.937}$ & $_{-0.815}^{+3.966}$ \\ 
NGC2099&  2099-WD2          & $19900\pm 900$ & $8.11\pm 0.16$ & $ 0.693\pm 0.098 $ & $ 7.961\pm 0.165 $ &  $ 3.120 $ & $_{-0.071}^{+0.105}$ & $_{-0.162}^{+0.201}$ \\ 
(M37)  &  2099-WD3          & $18300\pm 900$ & $8.23\pm 0.21$ & $ 0.764\pm 0.128 $ & $ 8.190\pm 0.190 $ &  $ 3.300 $ & $_{-0.162}^{+0.282}$ & $_{-0.197}^{+0.258}$ \\ 
       &  2099-WD4          & $16900\pm 1100$& $8.40\pm 0.26$ & $ 0.870\pm 0.153 $ & $ 8.428\pm 0.217 $ &  $ 3.773 $ & $_{-0.435}^{+2.146}$ & $_{-0.324}^{+0.495}$ \\ 
       &  2099-WD5          & $18300\pm 1000$& $8.33\pm 0.22$ & $ 0.828\pm 0.131 $ & $ 8.271\pm 0.188 $ &  $ 3.406 $ & $_{-0.200}^{+0.453}$ & $_{-0.219}^{+0.305}$ \\ 
       &  2099-WD7          & $17800\pm 1400$& $8.42\pm 0.32$ & $ 0.883\pm 0.196 $ & $ 8.378\pm 0.289 $ &  $ 3.623 $ & $_{-0.406}^{+2.733}$ & $_{-0.278}^{+0.405}$ \\ 
       &  2099-WD9          & $15300\pm 400$ & $8.00\pm 0.08$ & $ 0.613\pm 0.053 $ & $ 8.268\pm 0.067 $ &  $ 3.401 $ & $_{-0.087}^{+0.122}$ & $_{-0.218}^{+0.302}$ \\ 
       &  2099-WD10         & $19300\pm 400$ & $8.20\pm 0.07$ & $ 0.748\pm 0.043 $ & $ 8.090\pm 0.068 $ &  $ 3.205 $ & $_{-0.044}^{+0.065}$ & $_{-0.178}^{+0.225}$ \\ 
       &  2099-WD11         & $23000\pm 600$ & $8.54\pm 0.10$ & $ 0.961\pm 0.056 $ & $ 8.149\pm 0.095 $ &  $ 3.257 $ & $_{-0.079}^{+0.104}$ & $_{-0.188}^{+0.242}$ \\ 
       &  2099-WD12         & $13300\pm 1000$& $7.91\pm 0.12$ & $ 0.550\pm 0.071 $ & $ 8.394\pm 0.126 $ &  $ 3.665 $ & $_{-0.252}^{+0.575}$ & $_{-0.290}^{+0.429}$ \\ 
       &  2099-WD13         & $18200\pm 400$ & $8.27\pm 0.08$ & $ 0.789\pm 0.052 $ & $ 8.229\pm 0.074 $ &  $ 3.348 $ & $_{-0.085}^{+0.110}$ & $_{-0.207}^{+0.278}$ \\ 
       &  2099-WD14         & $11400\pm 200$ & $7.73\pm 0.16$ & $ 0.450\pm 0.079 $ & $ 8.487\pm 0.085 $ &  $ 4.011 $ & $_{-0.326}^{+0.623}$ & $_{-0.400}^{+0.674}$ \\ 
       &  2099-WD16         & $13100\pm 500$ & $8.34\pm 0.10$ & $ 0.825\pm 0.062 $ & $ 8.689\pm 0.082 $ &  $ 9.136 $ & $_{-3.864}^{+9.990}$ & $_{-3.737}^{+9.990}$ \\ 
NGC2168&  NGC2168 LAWDS 1   & $32400\pm 512$ & $8.40\pm 0.125$& $ 0.894\pm 0.074 $ & $ 7.371\pm 0.220 $ &  $ 4.839 $ & $_{-0.163}^{+0.303}$ & $_{-0.750}^{+1.642}$ \\ 
(M35)  &  NGC2168 LAWDS 2   & $32700\pm 603$ & $8.34\pm 0.080$& $ 0.859\pm 0.048 $ & $ 7.232\pm 0.148 $ &  $ 4.735 $ & $_{-0.077}^{+0.123}$ & $_{-0.704}^{+1.464}$ \\ 
       &  NGC2168 LAWDS 5   & $52600\pm 1160$& $8.24\pm 0.095$& $ 0.828\pm 0.053 $ & $ 6.223\pm 0.047 $ &  $ 4.508 $ & $_{-0.003}^{+0.002}$ & $_{-0.602}^{+1.164}$ \\ 
       &  NGC2168 LAWDS 6   & $55200\pm 897$ & $8.28\pm 0.065$& $ 0.852\pm 0.038 $ & $ 6.130\pm 0.048 $ &  $ 4.504 $ & $_{-0.002}^{+0.002}$ & $_{-0.601}^{+1.159}$ \\ 
       &  NGC2168 LAWDS 15  & $29900\pm 318$ & $8.48\pm 0.060$& $ 0.934\pm 0.034 $ & $ 7.652\pm 0.093 $ &  $ 5.239 $ & $_{-0.192}^{+0.238}$ & $_{-0.932}^{+2.618}$ \\ 
       &  NGC2168 LAWDS 27  & $30500\pm 397$ & $8.52\pm 0.061$& $ 0.957\pm 0.035 $ & $ 7.676\pm 0.092 $ &  $ 5.298 $ & $_{-0.217}^{+0.260}$ & $_{-0.963}^{+2.793}$ \\ 
Sirius &  Sirius B          & $25000\pm 200$ & $8.60\pm 0.04$ & $ 1.000\pm 0.020 $ & $ 8.091\pm 0.020 $ &  $ 5.056 $ & $_{-0.171}^{+0.262}$ & $_{-0.213}^{+0.273}$ \\
\hline
\end{tabular}
\end{minipage}
\end{table*}

\section{Initial to final mass relationship}

The data in Figure 1 span the initial mass range $\sim 2.5-6.5\msun$ and thuse
do not provide information at the low and high mass ends of single stars
evolving into white dwarfs.  We note that the IFMR of most clusters appears to
have an intrinsic spread, that is, there is a range of initial masses which
gives rise to a given final mass. If this spread is real (that is, if we can
dismiss the possibility that the spread is not caused, for example, by non
membership of the cluster, effects introduced by binaries, or by the
underestimation of errors, as could be the case for NGC2099) then the IFMR
should be characterised by a mean relationship and a spread about this
mean. This spread could in principle also be modelled in future when enough
data become available.
\begin{figure}
\begin{center}
\hspace{0.1in}
\epsfxsize=0.98\columnwidth
\epsfbox[18 145 573 592]{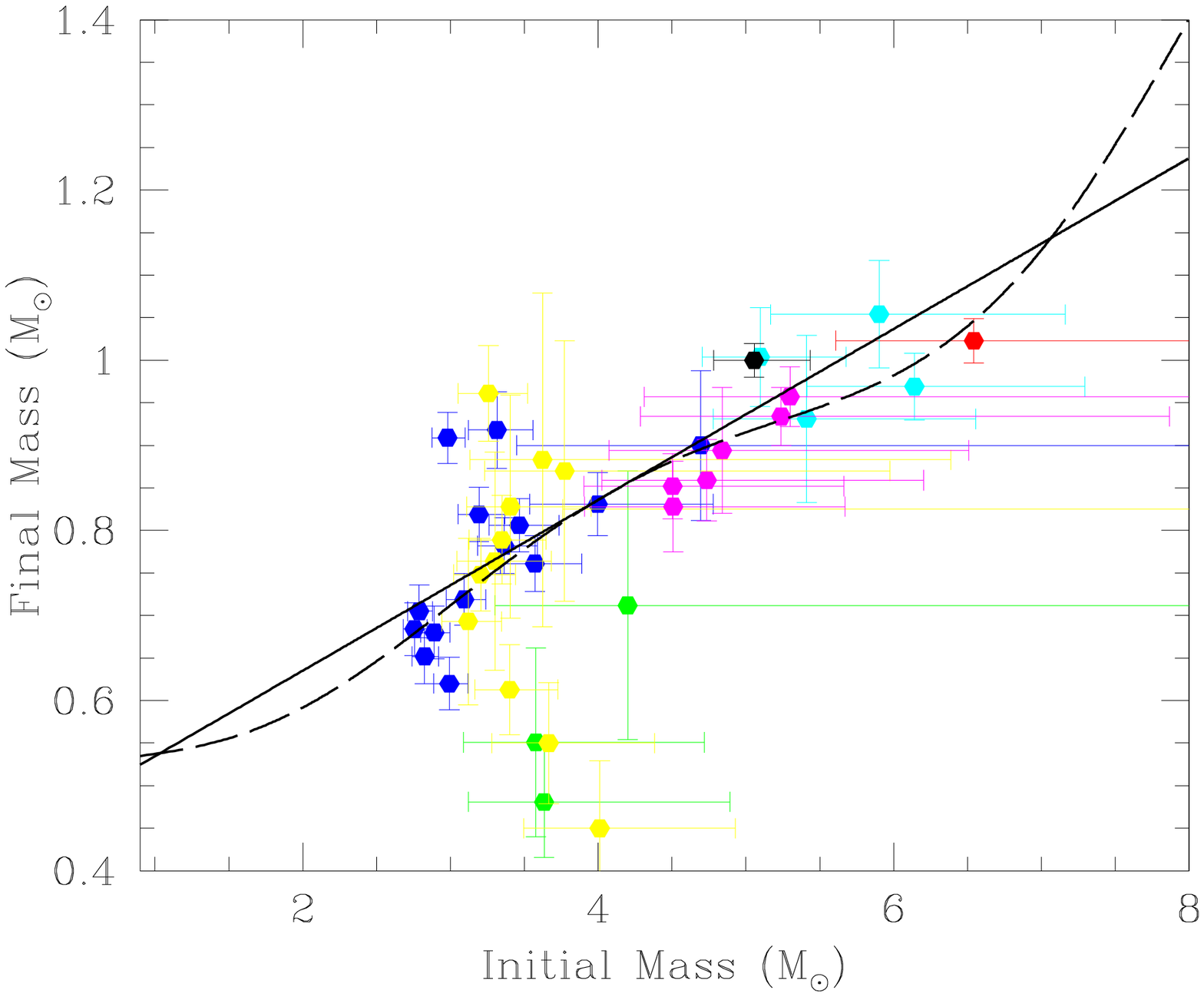}
\caption{The ensemble of data for the seven open clusters defining the initial
  to final mass relation. The error bars correspond to the best estimates of
  the cluster ages. Red: Pleiades; Cyan: NGC2516; Magenta: NGC2168 (M35);
  Blue: Praesepe and Hyades; Green: NGC3532; Yellow: NGC2099 (M37); Black:
  Sirius. The solid line is the best linear fit to the data and the dashed
  line is the IFMR with curvature (see text).}
\end{center}
\end{figure}

Given the various uncertainties involved, we cannot justify the use of any but
a linear relationship to model the cluster data. When allowance is made for
the uncertainties in the cluster ages, the error bars increase significantly
particularly at the high mass end, and the weight is shifted from the
high-mass end to the low-mass end of this diagram. Given the scatter in this
part of the diagram, a least squares linear fit to the data leads to a
meaningless result.  We have circumvented this problem by introducing an
anchor point at ($M_i=1.1\pm 0.1\msun, M_f= 0.55\pm 0.01\msun$)
(c.f. Weidemann 2000) which we justify as a requirement to reproduce the well
established peak in the field white dwarf mass distribution at $M_f=0.57\msun$
(see next section). With this anchor point, we obtain the best fit linear
relationship to be
\[
M_f=(0.10038\pm 0.00518)~M_i+0.43443\pm 0.01467
\]

We note that if we adopt the best estimate for the cluster ages, a best fit
linear relationship can be obtained from the cluster data alone without an
anchor point at the low-mass end. The linear relationship that we obtain with
this assumption is indistinguishable from the above equation within the given
errors.

\section{Population synthesis}

We show in Figure 2 the expected mass distribution of DA white dwarfs deduced from the PG survey
corrected to an equivalent volume limited sample (Liebert et al. 2005a). 
\begin{figure}
\begin{center}
\hspace{0.1in}
\epsfxsize=0.98\columnwidth
\epsfbox[47 165 575 696]{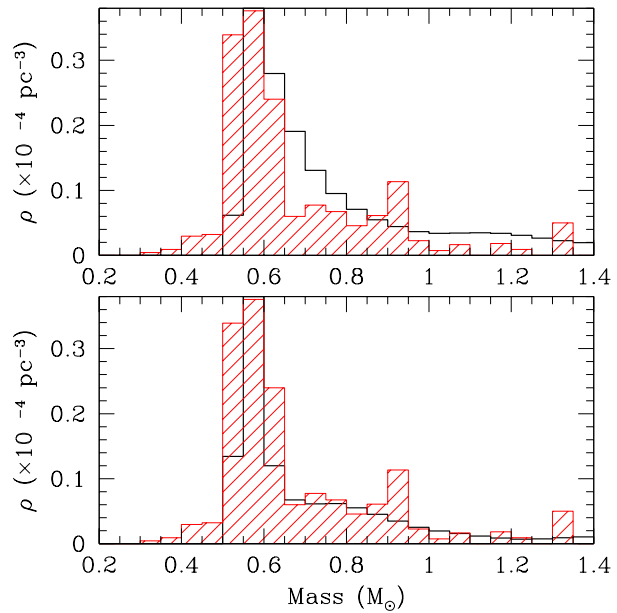}
\caption{Weighted observed mass distribution of field white dwarfs in the
  Palomar-Green survey (Liebert et al. 2005a, shaded histogram) for stars
  hotter than 13,000~K with the best linear IFMR fit model (top panel) and the
  IFMR with curvature (bottom panel). White dwarfs with masses less than
  $0.5\msun$ are believed to be helium white dwarfs resulting from close
  binary evolution and are not modelled in this paper.}
\end{center}
\end{figure}

Carbon-Oxygen white dwarfs resulting from single star evolution are not
expected to have a mass less than $0.5\msun$.  The white dwarfs with lower
masses in the PG survey must have interiors composed of helium and result from
binary evolution during which the envelopes have been stripped off before
helium ignition. We do not consider such stars in our population synthesis
calculations.

We have carried out calculations of the expected mass distribution of field
white dwarfs for different assumptions on the IFMR.  We follow closely the
method described by WF05 and we refer to that paper for details.

In essence, we calculate the number of white dwarfs per unit volume per unit
final mass interval with luminosities in the range $L_1$ to $L_2$ by the
integral
\[
\frac{dN_{WD}(M_f,L_1,L_2)}{dM_f}= \int_{\tau_c}^
{\tau_d}\frac{\psi(t)}{H(\tau_{disc}- 
t,M_i)}\phi(M_i)\frac{dM_i}{dM_f}dt
\]
with:
\[
\tau_{c}=\tau_{disc}- \tau_{evol}(M_i)-\tau_{wd}(M_f(M_i),L_2)
\]
\[
\tau_{d}=\tau_{disc}- \tau_{evol}(M_i)-\tau_{wd}(M_f(M_i),L_1)
\] 
and where
$\phi(M_i)$ is the initial mass function, $\psi(t)$ is the star formation rate
per unit area of the galactic disc, $\tau_{\mathit{disc}}$ is the age of the
disc, $\tau_{\mathit{evol}}$ is the time of evolution of a Main Sequence star
of mass $M_i$ to the beginning of the white dwarf phase (we assume that all
white dwarfs are born at luminosity $10^2 \lsun$), $t_{wd}$ is the white dwarf
cooling time to a luminosity $L$ and $H(\tau_{disc}- t,M_i)$ is a possibly
time-dependent scale factor which allows for the inflation of the galactic
disc. The parameters used here are as in WF05.

At the basis of our calculations is the assumed form for the IFMR. We consider
only those stars with $L_1>2.5\times^{-3}\lsun$, which corresponds to the
luminosity of a white dwarf of $M=0.8\msun$ at $\Teff=13,000$ (Wood 1995).  We
consider this to be appropriate in an average sense for comparison of our
results with the data histogram obtained from the PG survey. We show in Figure
2 the results that we obtain when we use the linear IFMR overlapped to
the data. The theoretical predictions are in overall agreement with the
observations for $M>0.5\msun$ although the shape is not reproduced in detail.
The model predicts $\sim 29$ percent of white dwarfs of all types to have
masses greater than $0.8\msun$. This value is similar to the $\sim 30$ percent
in the PG survey (Liebert et al 2005a).  The inclusion of non-DA spectral
types could also have a small ($\sim 10$ percent) effect on the relative
numbers and percentages.

The observed white dwarf mass distribution appears to show a deficit of stars
relative to the model in the mass range $0.65 - 0.8\msun$ and an excess in the
mass range $1.0- 1.4\msun$. Rapid changes in the slope of the IFMR are
reflected in the white dwarf mass distribution so that these deviations can be
explained if we move away from a linear IFMR. A well determined white dwarf
mass distribution can in principle be inverted to establish the appropriate
IFMR. We demonstrate that the introduction of curvature into the IFMR in the
appropriate mass ranges does indeed result in better overall agreement with
the observed white dwarf mass distribution. The relationship that we have used
is given by
\begin{align*}
M_f =&-0.00012336M_i^6+ 0.003160M_i^5-0.02960M_i^4 \\
& +0.12350M_i^3-0.21550M_i^2+0.19022M_i+0.46575
\end{align*}
With this relationship, there are changes in slope in the white dwarf mass
distribution at $0.65\msun$ and at $0.95\msun$ as indicated by the PG
survey. The cluster data may provide independent evidence for the presence of
such a curvature when a larger and better refined dataset becomes available
through future cluster observations.

The interpretation of the mass distribution of field white dwarfs is
complicated by the fact that there will be a component that has resulted from
close binary evolution. We expect in particular that white dwarf mergers will
play a dominant role at the very high mass end ($M\gsimeq 1\msun$) of this
distribution.

\section{Discussion and Conclusions}

The initial-final mass distribution can be derived empirically by studying
white dwarfs in open clusters, or by modelling the observed mass distribution
of white dwarfs. Uncertainties in cluster ages, metallicity, and in
atmospheric model fitting to spectral data result in larger error bars being
associated with many of the cluster data points. In this paper, we have
brought together all presently available data on the IFMR and made the
simplest hypothesis that the the IFMR can be modelled by a mean relationship
about which there is an intrinsic scatter.  We have taken the point of view
that a linear fit is all that can be warranted by the presently available
cluster data and that the data set is not sufficiently extensive to warrant
the modelling of the scatter.

The IFMR can be used to predict the white dwarf mass distribution. Here, the
analysis depends on the input to the population synthesis calculations, such
as the well established initial mass function, and the age of the galactic
disc. The peak in the observed mass distribution is determined by the lowest
mass main sequence stars that evolve into white dwarfs, while the high-mass
tail is determined by the profile of the IFMR for high-mass progenitors. We
have shown while the best fit linear relationship to the ensemble of cluster
data with a suitably chosen anchor point at the low mass end predicts a white
dwarf mass distribution that is in general agreement with observations, a
relationship with curvature improves the quality of the fit to the white dwarf
mass distribution.  For both the linear and curved IFMRs, some $\sim 28$
percent of the stars in the white dwarf mass distribution has masses above
$\sim 0.8\msun$. It is therefore not necessary to postulate the presence of a
dominant component in the high-mass tail that arises from mergers (e.g.
Liebert et al. 2005a), even though we may expect more mergers in high galactic
latitude surveys such as the PG, compared to, for example, the EUV survey
which roughly tracks the B-type stars of Gould's belt.

An interesting corollary to our present study relates to the properties of the
HFMWDs. It has been know for quite some time that the HFMWDs had masses that
were, on the average, higher than for the non-magnetic white dwarfs.  Early
studies suggested that $\sim 4$ percent of the white dwarfs were magnetic, and
such a percentage could be explained by assuming that their progenitors were
the known magnetic Ap and Bp stars which comprised $\sim 10$ percent of all A
and B stars. However, when appropriate corrections were made to allow for the
fact that most surveys were magnitude limited, the predicted proportion of
magnetic white dwarfs increased by a factor of $\sim 2-3$ (Kawka 2004,
Liebert et al. 2003). WF05 argued that if one dismissed the possibility of a
IFMR that is affected by magnetic fields, both the observed mass distribution,
and the observed space density of the HFMWDs could be reproduced if in
addition to the currently recognised group of magnetic Ap and Bp stars, $40$
percent of stars with masses in the range $\sim 4.5 - 10\msun$ also evolved
into HFMWDs.

This conclusion is now seen to be consistent with the results of the present
study where $40-50$ percent of the $\sim 28$ percent of stars in the
high-mass tail of the white dwarf mass distribution provides the observed
$\sim 10$ percent of the HFMWDs.

\section*{Acknowledgements}
The authors are indebted to Pierre Bergeron for providing access to his white
dwarf models, Ted von Hippel for valuable discussions and the Referee, Detlev
Koester, for his helpful comments.

\section{References}

\reference Boesgaard A.M., Friel E.D., 1990, ApJ, 351, 467

\reference Cayrel R., Cayrel de Strobel G., Campbell B., 1985 A\&A, 146 249

\reference Claver C.F., Liebert J., Bergeron P., Koester D., 2001, ApJ, 563, 
987

\reference Dobbie P.D., Pinfield D.J., Napiwotzki R., Hambly N.C.,
Burleigh M.R., Barstow M.A., Jameson R.F., Hubeny I., 2004, 355, L39

\reference Fontaine G., Brassard P., Bergeron P., 2001, PASP, 113, 409

\reference Girardi L., Bressan A., Bertelli G., Chiosi C., 2000,
A\&A. (Suppl. Ser.), 141, 371

\reference Girardi L., Bertelli G., Bressan A., Chiosi C., Groenewegen
M.A.T., Marigo P., Salasnich B., Weiss A., 2002, A\&A, 391, 195


\reference Kalirai J.S., Richer H.B. Reitzel D., Hansen B.M.S., Rich
R.M., Fahlman G.G., Gibson B.K., von Hippel T., 2005, ApJL, 618, L123

\reference Kawka A., 2004, Ph.D. (Thesis), `A study of white dwarfs in the
solar neighbourhood', Murdoch University

\reference King J.R., Hiltgen D.R., 1996, AJ 112, 2650

\reference King J.R., Soderblow D.R., 2000, ApJ, 533, 944

\reference Koester D., Reimers D., 1993, A\&A, 275, 479

\reference Koester D., Reimers D., 1996, A\&A, 313, 810

\reference Koester D., Weidemann V., 1980, A\&A, 81, 145

\reference Liebert J., Bergeron P., Holberg J.B., 2003, AJ, 125, 348 

\reference Liebert J., Bergeron P., Holberg J.B., 2005a, ApJS, 156, 47

\reference Liebert J., Young P.A., Arnett D., Holberg J.B., and
Williams K.A., 2005b, ApJL, submitted

\reference Mazzei P., Pigatto L., 1989, A\&A, 193 148

\reference Meynet G., Mermilliod J.-C., Maeder A., 1993, A\&AS, 98, 477

\reference Perryman M.A.C., Brown A.G.A., Lebreton Y., Gomez A., Turon
C., de Strobel G.C., Mermilliod J.C., Robichon N., Kovalevsky J., Crifo
F., 1998, A\&A, 331, 81

\reference Sarajedini A., Brandt K., Grocholski A.J., Tiede G.P., 2004, AJ,
127, 991

\reference Stauffer J.R., Schulz G., Kirkpatrick J.D., 1998 ApJ, 499, 199.

\reference Steinhauer A., 2003, PhD thesis (Indiana U)  

\reference Sung H., Bessell M.S., 1999, MNRAS, 306, 361

\reference Sung H., Bessell M.S., Lee B.-W., Lee S.-G., 2002 AJ, 123, 290 

\reference Twarog B.A., Ashman K.M., Anthony-Twarog B.J., 1997, AJ, 114, 2556

\reference Ventura P., Zeppieri A., Mazzitelli I., D'Antona F., 1998, A\&A,
331, 1011

\reference von Hippel T., 2005, ApJ, 622, 565

\reference Weidemann, V., 2000, A\&A, 363, 647

\reference Wickramasinghe D.T., Ferrario L., 2005, MNRAS, 356, 1576 (WF05)

\reference Williams K.A., Bolte M., Koester D., 2004, ApJL, 615, L49

\reference Wood, M.A., 1995, LNP Vol.~443: White Dwarfs, 41

\label{lastpage}
\end{document}